\documentclass[useAMS,usenatbib]{mn2e}

\usepackage{graphicx}
\usepackage{txfonts}
\usepackage{times,epsfig} 
\usepackage[dvips]{color}
\usepackage[]{natbib}
\title[The puzzling radio source in the cool core cluster A 2626]{The puzzling radio source in the cool core cluster A 2626}
\author[M. Gitti]{M. Gitti$^{1,2,3}$\thanks{E-mail:
myriam.gitti@oabo.inaf.it} 
 \\ 
$^{1}$Physics and Astronomy Department, University of Bologna, via Ranzani 1, 40127 Bologna, Italy\\
$^{2}$INAF, Astronomical Observatory of Bologna, via Ranzani 1, 40127 Bologna, Italy\\
$^{3}$INAF, Istituto di Radioastronomia di Bologna, via Gobetti 101, I-40129 Bologna, Italy}

\begin{document}

\date{Accepted 2013 August 14}

\pagerange{\pageref{firstpage}--\pageref{lastpage}} \pubyear{2013}

\maketitle

\label{firstpage}

\begin{abstract}
  We report on new VLA radio observations performed at 1.4 GHz and 4.8
  GHz with unprecedented sensitivity and angular resolution ($\sim$1
  arcsec) of the cool core cluster A 2626, which is known to possess a
  radio mini-halo at its center.  The most unusual features of A 2626
  are two elongated radio features detected in previous observations
  to the north and south, having morphologies not common to the
  typical jet-lobe structures in cool cores.  In our new sensitive
  images the two elongated features appears clearly as bright radio
  arcs, and we discover the presence of a new arc to the west. These
  radio arcs are not detected at 4.8 GHz, implying a steep ($\alpha
  >1$) spectrum, and their origin is puzzling.  After subtracting the
  flux density contributed by these discrete features from the total
  flux measured at low resolution, we estimate a residual
  {\color{black}$18.0\pm1.8$} mJy flux density of diffuse radio
  emission at 1.4 GHz. We therefore confirm the detection of diffuse
  radio emission, which appears distinct from the discrete radio arcs
  embedded in it. Although its radio power is lower ($1.4 \times
  10^{23}$ W Hz$^{-1}$) than previously known, the diffuse emission
  may still be classified as a radio mini-halo.

\end{abstract}



\vspace{-0.1in}
\section{Introduction}

The central dominant (cD) galaxies of cool core clusters have a high
incidence of radio activity, showing the presence of central FR-I
radiogalaxies in 70\% of the cases \citep{Burns_1990, Best_2007,
  Mittal_2009}.  Their behaviour differs from that of quasar: in many
low-accretion-rate AGNs almost all the released energy is channelled
into jets because the density of the gas surrounding the black hole is
not high enough for an efficient radiation
\citep[e.g.,][]{Churazov_2005}. In fact, the importance of these
objects has been underestimated for a long time due to their poor
optical luminosity, and began to emerge after the discovery, with the
X-ray satellite {\it ROSAT}, of deficits in the X-ray emission of the
Perseus and Cygnus A clusters which are spatially coincident with
regions of enhanced synchrotron emission \citep{Bohringer_1993,
  Carilli_1994}.  With the advent of the new high-resolution X-ray
observations performed with {\it Chandra} and {\it XMM-Newton}, it
became clear that the central radio sources have a profound,
persistent effect on the ICM -- the central hot gas in many cool core
systems is not smoothly distributed, but shows instead 'holes' on
scales often approximately coincident with lobes of extended radio
emission.  The most typical configuration is for jets from the central
dominant elliptical of a cluster to extend outwards in a bipolar flow,
inflating lobes of radio-emitting plasma (radio {\it
  'bubbles'}). These lobes push aside the X-ray emitting gas of the
cluster atmosphere, thus excavating depressions in the ICM which are
detectable as apparent {\it 'cavities'} in the X-ray images.  Radio
galaxies have thus been identified as a primary source of feedback in
the hot atmospheres of galaxy clusters and groups \citep[for recent
reviews see][and references therein]{Gitti_2012,
  McNamara-Nulsen_2012}.

In some cases, the powerful radio galaxies at the center of cool core
clusters are surrounded by diffuse radio emission on scales $\sim$
$200-500$ kpc having steep radio spectra ($\alpha > 1 ; S_{\nu}
\propto \nu^{-\alpha}$). These radio sources, generally referred to as
{\it `radio mini-halos'}, are synchrotron emission from GeV
electrons diffusing through $\mu$G magnetic fields.  Although the
central radio galaxy is the obvious candidate for the injection of the
population of relativistic electrons, mini-halos do appear quite
different from the extended lobes maintained by AGN, therefore their
radio emission proves that magnetic fields permeate the ICM and at the
same time may be indicative of the presence of diffuse relativistic
electrons.  In particular, due to the fact that the radiative lifetime
of radio-emitting electrons ($\sim 10^8$ yr) is much shorter than any
reasonable transport time over the cluster scale, the relativistic
electrons responsible for the extended radio emission from mini-halos
need be continuously re-energized by various mechanisms associated
with turbulence in the ICM (reaccelerated {\it primary} electrons), or
freshly injected on a cluster-wide scale (e.g. as a result of the
decay of charged pions produced in hadronic collisions, {\it
  secondary} electrons).  \citet{Gitti_2002} developed a theoretical
model which accounts for the origin of radio mini--halos as related to
electron reacceleration by magnetohydrodynamic (MHD) turbulence,
which is amplified by compression in the cool cores.  In this model,
the necessary energetics to power radio mini-halos is supplied by the
cooling flow process itself, through the compressional work done on
the ICM and the frozen-in magnetic field.  
Although secondary electron models have been proposed to explain the
presence of their persistent, diffuse radio emission on large-scale in
the ICM \citep[e.g.,][]{Pfrommer-Ensslin_2004, Keshet-Loeb_2010}, the
observed trend between the radio power of mini-halos and the maximum
power of cooling flows \citep{Gitti_2004, Gitti_2012} has given
support to a primary origin of the relativistic electrons radiating in
radio mini-halos, favored also by the successful, detailed application
of the \cite{Gitti_2002} model to two cool core clusters
\citep[Perseus and A 2626,][]{Gitti_2004} and by recent statistical
studies \citep{Cassano_2008}.  However, the origin of the turbulence
necessary to trigger the electron reacceleration is still debated.
The signatures of minor dynamical activity have recently been detected
in some mini-halo clusters, thus suggesting that additional or
alternative turbulent energy for the reacceleration may be provided by
minor mergers \citep{Gitti_2007b} and related gas sloshing mechanism
in cool core clusters \citep{Mazzotta-Giacintucci_2008, ZuHone_2013}.

Radio mini-halos are rare, with only about a dozen objects known so
far \citep{Feretti_2012}.  The criteria initially adopted by
\citet{Gitti_2004} to select the first sample of mini-halo clusters,
i.e. the presence of both a cool core and a diffuse, amorphous radio
emission with no direct association with the central radio source,
having size comparable to that of the cooling region, are now
typically used to identify mini-halos.  However, these criteria are
somehow arbitrary (in particular, total size, morphology, presence of
cool core) and some identifications are still controversial.
Furthermore, we stress that the classification of a radio source as a
mini-halo is not trivial: their detection is complicated by the fact
that the diffuse, low surface brightness emission needs to be
disentangled from the strong radio emission of the central radio
galaxy and of other discrete sources.

A 2626 is a low-redshift {\color{black}
  \citep[z=0.0553,][]{Struble-Rood_1999}}, regular, relatively poor
Abell cluster \citep{Mohr_1996}, with a double-nuclei cD elliptical
galaxy (IC 5338) showing extended strong emission lines
\citep{Johnstone_1987}.  Early observations with {\it Einstein} and
{\it ROSAT}, then confirmed by {\it Chandra} {\color{black} and {\it
    XMM--Newton}}, indicate that it hosts a moderate cooling flow
\citep{White_1991, Rizza_2000, Wong_2008}.  Early, low resolution
radio surveys showed that this cluster contains a central radio source
exhibiting a compact unresolved core and a diffuse structure with very
steep radio spectrum \citep{Slee-Siegman_1983, Roland_1985,
  Burns_1990}.  The compact radio component is associated with the
southwest nucleus of the cD galaxy IC 5338 \citep{Owen_1995}.  More
recently, high resolution VLA images of \citet{Gitti_2004} highlighted
the unusual properties of the A 2626 radio source.  These authors
found that at 1.4 GHz the central component consists of an unresolved
core plus a small jet-like feature pointing to the south-western
direction.  The extended emission, that at lower resolution appears as
a diffuse diamond--shaped emission detected up to $\sim 1'$ from the
cluster center, is resolved out and two elongated parallel features of
similar brightness and extent are visible.  The total flux density of
these `bar' structures is $\sim 6.6$ mJy, contributing to $\sim$ 20\%
of the flux of the diffuse radio emission detected at low resolution.
Such symmetric, elongated features are imaged also by low resolution
observations at 330 MHz, with a total flux density (including the
diffuse emission) of $\sim 1$ Jy, whereas no radio emission is
detected at the location of the core.
\citet{Gitti_2004} argue that the two unusual elongated features are
distinct from and embedded in the diffuse extended radio emission,
which they classified as a radio mini--halo and successfully modeled
as radio emission from relativistic electrons reaccelerated by MHD
turbulence in the cooling core region. On the other hand, the origin
and nature of the two radio bars is not clear as they are not
associated to any X-ray cavities \citep{Wong_2008}.

In this paper we present new high-resolution VLA data of the central
radio source in A 2626, whose morphology much complex than that of the
standard X-ray radio bubbles seen in other cool core clusters
represents a challenge to models for the ICM - radio source
interaction.  With $H_0 = 70 \mbox{ km s}^{-1} \mbox{ Mpc}^{-1}$, and
$\Omega_M=1-\Omega_{\Lambda}=0.3$, the luminosity distance of A 2626
is {\color{black} 246.8} Mpc and 1 arcsec corresponds to 1.1 kpc.


\vspace{-0.15in}
\section{Observations and Data Reduction}

\begin{table}
\begin{center}
\caption{New VLA data analyzed in this paper (project code: AG795, PI: M. Gitti)}
\begin{tabular}{lccccc}
\hline
\hline
Obs. Date   & Band &Frequency & Bandwidth & Array & TOS \\
            &      & (MHz)    & (MHz)     & ~     & (h) \\
\hline
~&~&~&~&~&~\\
08-Nov-21 & C & 4885.1/4835.1 & 50.0 & A & 3.5 \\
09-Apr-30 & C & 4885.1/4835.1 & 50.0 & B & 3.5 \\
08-Nov-22 & L & 1464.9/1385.1 & 50.0 & A & 5.0 \\
09-May-04 & L & 1464.9/1385.1 & 50.0 & B & 7.0 \\
\hline
\label{vladata.tab}
\end{tabular}
\end{center}
\vspace{-0.2in}
\end{table}

We performed new Very Large Array\footnote{The Very Large Array (VLA)
  is a facility of the National Radio Astronomy Observatory (NRAO).
  The NRAO is a facility of the National Science Foundation, operated
  under cooperative agreement by Associated Universities, Inc.}
observations of the radio source A 2626 at 1.4 GHz and 4.8 GHz in A-
and B-configuration (see Table \ref{vladata.tab} for details regarding
these observations).  In all observations the source 3C 48
(0137$+$331) was used as the primary flux density calibrator, while
the sources 0016$-$002 and 3C 138 (0521$+$166) were used as secondary
phase and polarization calibrators, respectively.  Data reduction was
done using the NRAO AIPS (Astronomical Image Processing System)
package, version 31DEC13.  Accurate editing of the {\it (u,v)} data
was applied with the task {\ttfamily TVFLG} to identify and remove bad
visibility points.  Images were produced by following the standard
procedures: Calibration, Fourier-Transform, Clean and Restore.
Self-calibration with the task {\ttfamily CALIB} was applied to remove
residual phase variations.

The four datasets, taken at different frequencies and with different
configurations, were reduced separately in order to optimize the
accuracy of data editing and (self-)calibration of each single
observation.  In order to fully exploit the relative advantages in
terms of angular resolution and sensitivity of the two VLA
configurations used for the observations, we then combined together
(with the task {\ttfamily DBCON}) the resulting \textit{(u,v)} data in
A- and B- configuration taken at the same frequency. On these combined
A$+$B datasets, one at 1.4 GHz and one at 4.8 GHz, we also performed a
few more iterations of self-calibration to improve the dynamic range
of the images.  We produced images in the Stokes parameters I, Q and U
at different resolutions by specifying appropriate values of the
parameters {\ttfamily UVTAPER} and {\ttfamily ROBUST} in the AIPS task
{\ttfamily IMAGR}. The images of the polarized intensity, the
fractional polarization and the position angle of polarization were
derived from the I, Q and U images. The final maps show the contours
of the total intensity.

For the only purpose of deriving the total flux density of the radio source,
including the diffuse emission, we also analyzed old 1.4 GHz data
obtained with the VLA in C-configuration (Proj. code AM735, PI:
T. Markovic). In this 3.5 h archival observation, performed in
spectral-line mode, the source 3C 48 was used as the primary flux
density calibrator, while the source 0204$+$152 was used as secondary
phase calibrator.

Typical amplitude calibration errors are at 3\%, therefore we assume
this uncertainty on the flux density measurements. 


\vspace{-0.15in}
\section{Results}

\begin{figure}
\centering
\includegraphics[scale=0.45]{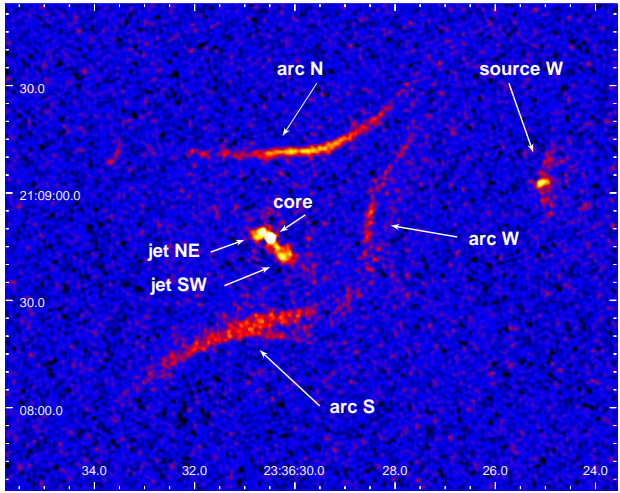}
\vspace{-0.2in}
\caption{ 1.4 GHz VLA map of A2626 at a resolution of $1''.2 \times
  1''.2$, obtained by setting the parameter {\ttfamily ROBUST=-5,
    UVTAPER=0}. The r.m.s. noise is 0.012 mJy/beam and the peak flux
  density is 12.9 mJy/beam.  
The arrows indicate the features discussed in the text.}
\label{map-1.4-r-5.fig}
\vspace{-0.1in}
\end{figure}

\begin{figure} 
\centering
\includegraphics[scale=0.35, angle=-90]{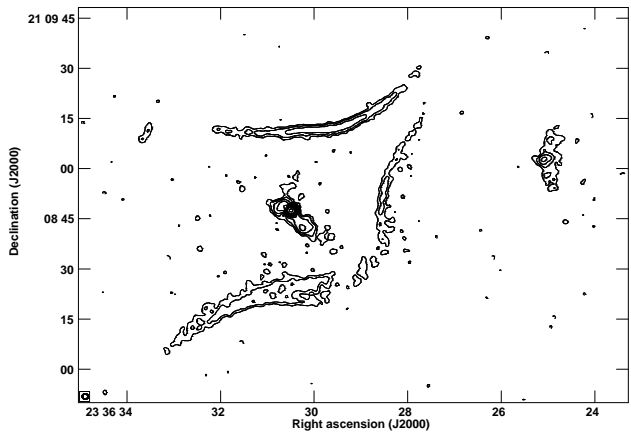}
\vspace{-0.2in}
\caption{
{\it Left:} 1.4 GHz VLA map of A2626 at a resolution of $1''.7 \times
  1''.6$ (the beam is shown in the lower-left corner), obtained by
  setting the parameter {\ttfamily ROBUST=0, UVTAPER=0}. The
  r.m.s. noise is 0.010 mJy/beam and the peak flux density is 13.1
  mJy/beam.  The contour levels are $-0.04$ (dashed), 0.04, 0.08, 0.16,
  0.32, 0.64, 1.28, 2.56, 5.12, 10.24 mJy/beam.  
}
\label{map-1.4-r0.fig}
\vspace{-0.1in}
\end{figure}

\begin{figure} 
\centering
\includegraphics[scale=0.45]{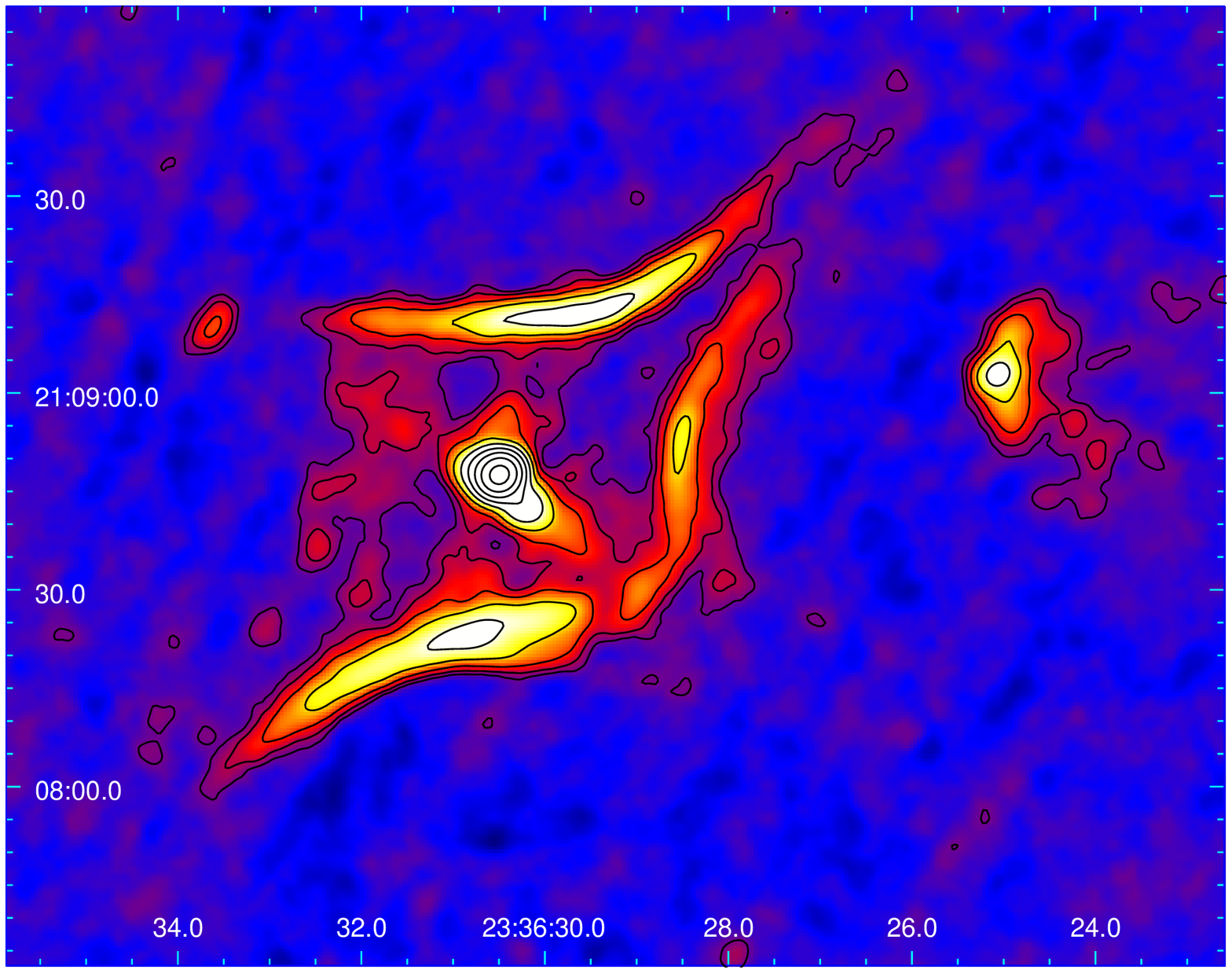}
\vspace{-0.2in}
\caption{ 1.4 GHz VLA map of A2626 at a resolution of $4''.4 \times
  3''.9$ (the beam is shown in the lower-left corner), obtained by
  setting the parameter {\ttfamily ROBUST=5, UVTAPER=90}. The
  r.m.s. noise is 0.014 mJy/beam and the peak flux density is 14.0
  mJy/beam.  The contour levels are the same as in Fig. \ref{map-1.4-r0.fig}. 
}
\label{map-1.4-r5.fig}
\vspace{-0.1in}
\end{figure}
 
Figure \ref{map-1.4-r-5.fig} shows the 1.4 GHz radio image of A2626 at
full resolution (restoring beam of $1''.2 \times 1''.2$), obtained
with pure uniform weighting by setting {\ttfamily ROBUST=-5}, whereas
Figure \ref{map-1.4-r0.fig} shows the contour map obtained by
tempering the uniform weights with {\ttfamily ROBUST=0} (restoring
beam of $1''.7 \times 1''.6$), which slightly improved the dynamic
range of the image.  With these high resolutions it is possible to
image the central source and to spot the presence of other discrete features
that can contribute to the total flux density and extended morphology detected
at lower resolution.
The nuclear source, located at RA (J2000): $23^{\rm h}$ $36^{\rm m}$
$30^{\rm s}.5$, Dec (J2000): $21^{\circ}$ $08'$ $47''.6$, is here
fully unveiled: it clearly shows two jets pointing to the
northeast-southwest direction, extending out from the unresolved core
to a distance of approximately 5 kpc.  The total flux density of the
core-jets structure is {\color{black} $17.7\pm0.5$ mJy}.
In these sensitive images the already known elongated radio features
stand out very clearly as bright radio arcs (arc N and arc S in Fig.
\ref{map-1.4-r-5.fig}). They are symmetrically located at $\sim25$ kpc
to the north and south of the core, showing total longitudinal
extensions of approximately 70 and 62 kpc, respectively.  Their
combined total flux density is {\color{black} $16.2\pm0.5$ mJy}, which is a
factor $\sim 2.5$ higher than previously estimated \citep{Gitti_2004}.
Furthermore, we identify a new feature which was not detected in the
previous observations: a faint {\color{black} ($3.1\pm0.1$ mJy)}, elongated
arc to the west of the core (arc W in Fig. \ref{map-1.4-r-5.fig}),
which extending for about 60 kpc appears to ideally connect the
western edges of the two radio arcs N-S.
No diffuse emission is seen at this high resolution and no significant
polarized flux is detected at 1.4 GHz at any resolution.

We produced an image at the lower resolution of $4''.4 \times 3''.9$,
obtained with tapered natural weighting by setting {\ttfamily
  ROBUST=5} and {\ttfamily UVTAPER=90} (Fig. \ref{map-1.4-r5.fig}), to
better map the extended structures.  Diffuse 1.4 GHz emission appears
clearly surrounding the nucleus in the region encompassed by the
features (arc N, arc S and arc W) discussed above.  In order to
estimate correctly the flux density of the diffuse emission, we
analyzed the archival C-array data and produced a 1.4 GHz map at a
resolution of $14''.4 \times 12''.6$ (not shown here).  The total
source A 2626 imaged with this low resolution has a well-known
diamond-like shape extending for approximately $2'.9 \times 2'.0$
($156 \times 135$ kpc), with a total flux density of {\color{black}
  $55.0\pm1.7$ mJy} (source W excluded), in agreement with previous
observations \citep{Ledlow-Owen_1995, Rizza_2000}. By subtracting the
emission contributed by the discrete features seen at high resolution
(core $+$ jets $+$ arcs = {\color{black} $37.0\pm0.6$ mJy}, see Table
\ref{risradio.tab}) from the total flux density of the C-array 1.4 GHz
emission, we thus measure a residual flux density of diffuse radio
emission of {\color{black} $18.0\pm1.8$ mJy}.

The more crude approach of measuring directly the flux density inside
the contours of the diffuse emission visible in the map in
Fig. \ref{map-1.4-r5.fig} would provide an estimate of $\sim$5
mJy. However, we note that this is certainly a lower limit as the
A$+$B array data misses the flux contribution of the short baselines
which are instead present in the C- array data, thus losing
sensitivity to diffuse radio structures.  We also attempted to derive
the flux of the discrete features from the C-array map directly,
getting a rough estimate of {\color{black} $\sim$40 mJy (unresolved central
  component $+$ arcs)}.  Considering the difficulty in separating the
emission from each component at low resolution, leading to big
uncertainties, this estimate should be considered not far from the
accurate one derived from our high resolution A$+$B array data.
The variety of methods discussed here demonstrates the complexity of
disentangling the relative contribution of the diffuse emission and of
the discrete sources to the total observed radio emission.

\begin{figure}
\centering
\includegraphics[scale=0.35, angle=-90]{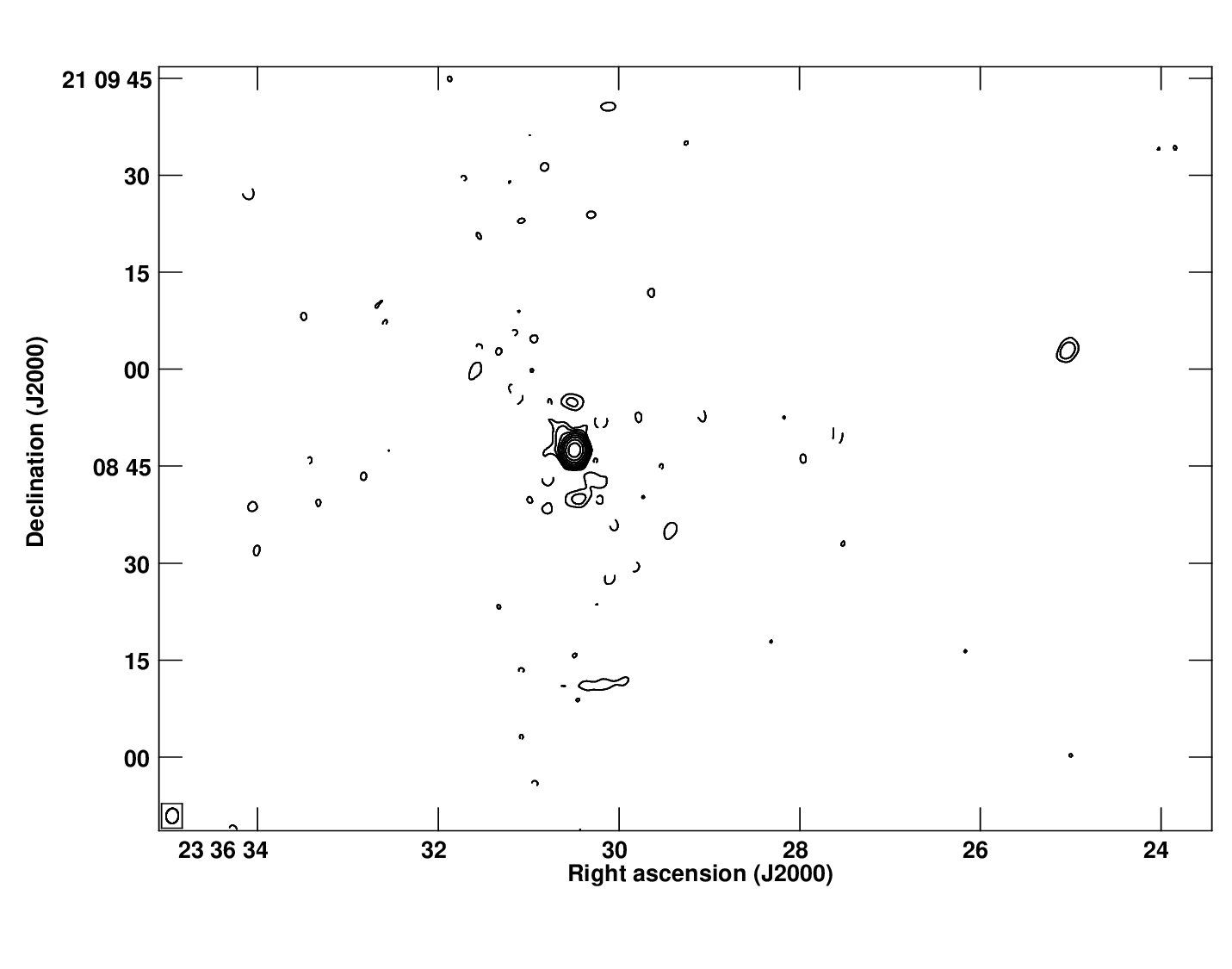}
\vspace{-0.2in}
\caption{ 4.8 GHz VLA map of A2626 at a resolution of $2''.3 \times
  2''.0$ (the beam is shown in the lower-left corner), obtained by
  setting the parameter {\ttfamily ROBUST=5, UVTAPER=90}. The
  r.m.s. noise is 0.012 mJy/beam and the peak flux density is 9.5
  mJy/beam.  The contour levels are the same as in
  Fig. \ref{map-1.4-r0.fig} and the map size in the same as
  Figs. \ref{map-1.4-r-5.fig}-\ref{map-1.4-r5.fig}.  }
\label{map-4.8-r5.fig}%
\vspace{-0.1in}
\end{figure}

Figure \ref{map-4.8-r5.fig} shows the 4.8 GHz radio image of A 2626 at
a resolution of $2''.3 \times 2''.0$, obtained with tapered natural
weighting by setting {\ttfamily ROBUST=5} and {\ttfamily UVTAPER=90}.
At this higher frequency only the central component is visible,
showing an unresolved core plus jet-like features pointing to the
{\color{black} northeast-southwest directions\footnote{{\color{black}Given the
    pattern of the dirty beam, the two 'blobs' visible to the north and
    south of the core are likely to be image artifacts}}. The total flux
  density is {\color{black}$9.8\pm0.3$} mJy (jet-like features
  included)}.  The core appears slightly polarized at a level of
$\sim$4\%.  The three discrete arcs seen at 1.4 GHz are not detected,
implying a steep spectral index (see Table \ref{risradio.tab}).

We finally note that a discrete source is visible to the west of A
2626 in all radio maps (source W in Fig. \ref{map-1.4-r-5.fig}),
located at RA(J2000):$23^{\rm h} 36^{\rm m} 25^{\rm s}.1$,
Dec(J2000):$21^{\circ} 09' 03''$, which is associated to the cluster
S0 galaxy IC 5337.  The morphology of this source suggests that it is
a head-tail radio galaxy. 

The radio results are summarized in Table \ref{risradio.tab}, where we
also report the 1.4 GHz monochromatic radio power.

\begin{table}
\vspace{-0.3in}
\caption{
\label{risradio.tab}
Summary of radio results {\color{black} for A 2626}.
The sizes are estimated from the maps at 1.4 GHz, and the flux densities are 
estimated inside the $3 \sigma$ contour level (for the core measurements, we 
performed a gaussian fit with the task {\ttfamily JMFIT}).
The superscripts $^{(1)}$, $^{(2)}$ and $^{(4)}$
indicate that the flux is measured from  Fig. \ref{map-1.4-r-5.fig}, 
Fig. \ref{map-1.4-r0.fig} and Fig. \ref{map-4.8-r5.fig}, respectively. 
When no emission is detected at 4.8 GHz, the spectral index is estimated assuming
an upper limit of $3 \sigma$. {\color{black} The 1.4 GHz monochromatic radio power
 is in unit of $10^{23}$W Hz$^{-1}$ and the associated error is at  3\%.}}
\centerline{
\begin{tabular}{lccccc}
\hline
\hline
Source          & Size             & S$_{1.4}$ & S$_{4.8}$ & $\alpha^{4.8}_{1.4}$           & $P_{1.4}$ \\
~               & ($\arcsec^2$)    & (mJy)    & (mJy)    & $(S_{\nu} \propto \nu^{-\alpha})$ & \\
\hline
~ & ~& ~& ~& ~& ~\\
[-3mm]
Core            & unresolved        &$13.5{\color{black}\pm0.4}$ $^{(1)}$      &  $9.6{\color{black}\pm0.3}$ $^{(4)}$     & ${\color{black}0.28\pm0.03}$  & 1.06\\ 
Jet NE          & ${\color{black}\sim} 5 \times 5$      & $1.9{\color{black}\pm0.1}$ $^{(1)}$      & {\color{black} $\sim0.1$} $^{(4)}$     & {\color{black} $\sim$ 2.4}  & 0.15\\ 
Jet SW          & ${\color{black}\sim} 6 \times 4$      & $2.3{\color{black}\pm0.1}$ $^{(1)}$      & {\color{black} $\sim0.1$} $^{(4)}$     & {\color{black} $\sim$ 2.6}  & 0.18\\
Arc N           & ${\color{black}\sim} 64 \times 4$     & $7.2{\color{black}\pm0.2}$ $^{(2)}$      & -- $^{(4)}$       & $>1.6$ & 0.57\\
Arc S           & ${\color{black}\sim} 56 \times 6$     & $9.0{\color{black}\pm0.3}$ $^{(2)}$      & -- $^{(4)}$       & $>1.0$ & 0.71\\
Arc W           & ${\color{black}\sim} 55 \times 3$     & $3.1{\color{black}\pm0.1}$ $^{(2)}$      & -- $^{(4)}$       & $>1.0$ & 0.24\\
\hline
\end{tabular}
}
\end{table}


\vspace{-0.15in}
\section{Discussion and Conclusions}

\subsection{Radio Arcs}

Since their discovery \citep{Gitti_2004}, the nature of the two
elongated radio features to the north and south directions (arc N and
arc S in Fig. \ref{map-1.4-r-5.fig}) has been a puzzle.  Their
symmetric positions on each side of the core suggests that they are
radio bubbles, but their thin, elongated shapes are unlike those
typically observed in cool core clusters.  They may represent radio
emitting plasma injected by the central source during an earlier
active phase, which has then propagated through the cool core region
in the form of buoyant subsonic plumes
\citep[e.g.,][]{Gull-Northover_1973, Churazov_2000,
  Bruggen-Kaiser_2001}.  However, such plumes are expected to have a
torus-like concavity \citep{Churazov_2000}, contrarily to the observed
shape in A 2626.  Furthermore, by comparison to radio lobes and
bubbles associated with other cool core dominant radio galaxies, one
might expect these to be regions of reduced X-ray emission surrounded
by bright rims.  In fact, previous {\it ROSAT} observations failed to
find strong X-ray deficit in the cool core of A 2626
\citep{Rizza_2000}.  Recently, this cluster has been studied in more
detail with {\it Chandra} and {\it XMM-Newton}, but yet no X-ray
cavities associated with the elongated radio features have been found
\citep{Wong_2008}.

\citet{Wong_2008} argue that the lack of obvious correlation between
the two symmetric parallel radio features and any structures in the
X-ray images may indicate that they are thin tubes parallel to the
plane of the sky, or that the radio plasma is mixed with the X-ray
gas, rather than displacing it.  These authors suggest that jet
precession might also provide an alternative explanation of the
peculiar radio morphology.  If two jets ejected towards the north and
south by the southwest cD nucleus are precessing about an axis which
is nearly perpendicular to the line-of-sight and are stopped at
approximately equal radii from the AGN (at a `working surface'), radio
emission may be produced by particle acceleration, thus originating
the elongated structures.  The impressive arc-like, symmetric
morphology of these features highlighted by the new high resolution
radio images may support this interpretation.

The discovery of a third elongated feature to the west (arc W in Fig.
\ref{map-1.4-r-5.fig}) further complicates the picture. It could
represent another radio bubble ejected in a different direction,
similarly to what observed in RBS 797 \citep{Gitti_2006}, but again
the absence of any correlation with the X-ray image and its ``wrong''
concavity, as well as the absence of its counterpart to the east,
disfavor this interpretation. In the model proposed by
\citet{Wong_2008}, it could also represent the result of particle
acceleration produced at a working surface by a third jet ejected to
the west. This interpretation would imply the existence of radio jets
emanating also from the northeast nucleus (separated by only $\sim 4$
kpc from the active southwest nucleus) of the cD galaxy IC 5338, which
however does not show a radio core.  

We stress that the radio arcs have an elongated morphology and steep
spectral index (see Table \ref{risradio.tab}).  These characteristics
are similar to those of cluster radio relics associated to particle
reacceleration due to shocks \citep{Feretti_2012}.  We also note that
each arc resembles the morphology of the large-scale structure of the
radio source 3C 338, which is disconnected from the core emission
having two-sided jets and has been interpreted as a relic structure
not related to the present nuclear activity \citep{Giovannini_1998}.
The presence of three such arcs in A 2626 makes the interpretation
even more puzzling.
Seen all together, the combined shape of the three arcs suggests that
they could trace the symmetric fronts of a single elliptical shock
originating from the core.  However, we notice again that their
concavity is not what one would expect from a shock propagating from
the center.  If they are cluster radio relic-like sources due to
reacceleration at a bow shock, the observed concavity suggests that
three distinct shocks should be propagating toward the center from
different directions.  However, the presence of such symmetric shocks
in the atmosphere of a relatively relaxed cluster seems very unlikely,
given also that {\it Chandra} observations failed to detect any
obvious X-ray edge ascribable to shock fronts. Furthermore, relic sources
are typically strongly polarized \citep{Feretti_2012}, whereas no
significant polarized flux is detected in our data (see Sect. 3).

It is also possible that the nature of the two bright radio arcs N-S
is different from that of the fainter arc W, which lacks an obvious
counterpart to the east and could be related to the merging of the
nearby S0 galaxy IC 5337 (see Fig. 16 of \citet{Wong_2008} for a
possible correlation with the {\it Chandra} hardness ratio
map). Therefore the complex morphology of the A 2626 radio source may
result from a combination of the scenarios presented above.

As it appears evident from this discussion, the new high-resolution
VLA data are not definitive to solve the puzzle of the
origin and nature of the radio arcs in A 2626.

\vspace{-0.15in}
\subsection{Diffuse radio emission}

With the new sensitive observations presented here, which improve by a
factor $\sim3$ the r.m.s. noise of the published maps
\citep{Gitti_2004}, we confirm the detection of diffuse 1.4 GHz
emission, having a radio power of $P_{1.4} = 1.4 \times 10^{23}$ W
Hz$^{-1}$.
As imaged at this high resolution ($\sim 4 ''$), the diffuse radio
emission appears to be confined in the region encompassed by the three
radio arcs, surrounding the nucleus, and shows a fragmented morphology
with total size $\sim 60$ kpc.

Although lower than previously estimated, the radio power of the
diffuse emission in A 2626 still follows the trend with the maximum
power of cooling flows, which is expected in the framework of the
model proposed by \citet{Gitti_2002, Gitti_2004}. This model links the
origin of radio mini-halos to radio emission from relativistic
electrons reaccelerated by Fermi mechanism associated with MHD
turbulence amplified by the compression of the magnetic field in the
cooling core region, thus supporting a direct connection between 
cool cores and radio mini-halos (see Fig. 3b of \citet{Gitti_2012} for
a recent version of the observed trend).

The relativistic electrons responsible for the diffuse emission may
also have been reaccelerated by turbulence generated by the sloshing
of the cool core gas \citep{Mazzotta-Giacintucci_2008,
  ZuHone_2013}. The presence of three symmetric sloshing sub-clumps in
the cluster atmosphere, although unlikely, might also induce local
electron (re)acceleration thus explaining the origin of the three
radio arcs discussed in Sect. 4.1.


\vspace{-0.1in}

\section*{Acknowledgments}

MG thanks G. Giovannini and D. Dallacasa for helpful advices during
the data reduction, and S. Giacintucci and K. Wong for providing
comments on the original manuscript. MG acknowledges the financial
contribution from contract ASI-I/009/10/0.

\vspace{-0.1in}

\bibliography{../bibliography-gitti}

\end{document}